\documentclass[12pt,a4paper]{article}
\usepackage[utf8]{inputenc}
\usepackage{amsmath}
\usepackage{amsfonts}
\usepackage{amssymb}
\usepackage{subcaption}
\usepackage{graphicx}
\usepackage{cite}
\usepackage{authblk}
\usepackage{hyperref}
\usepackage{kpfonts}
\usepackage{float}
\usepackage[toc]{appendix}
\title{Deviations from $\mu$-$\tau$ symmetry using $\Delta$(27) group on neutrino masses and mixings }
\date{}
\begin{document}
\author{Ph.Wilina\footnote{wilina.phd.phy@manipuruniv.ac.in} $^{1}$
 and N.Nimai Singh\footnote{nimai03@yahoo.com} $^{1,2}$\\
$^{1}$\small Department of Physics, Manipur University, Imphal-795003, India\\ 
$^{2}$\small Research Institute of Science and Technology,  Imphal-795003, India}
\maketitle
\begin{abstract}
Implication of neutrino mass model based on $\Delta$(27) discrete flavor symmetry, on parameters of neutrino oscillations, CP violation and effective neutrino masses is studied using type-I seesaw mechanism. The Standard Model particle content is extended by adding two additional Higgs doublets, three right-handed neutrinos and two scalar triplets under $\Delta$(27) symmetry predicting diagonal charged lepton mass matrix. This can generate the desired deviation from $\mu - \tau$  symmetry. The resulting neutrino oscillation parameters are well agreed with the latest global fit oscillation data. The sum of the three absolute neutrino mass eigenvalues, $\sum\limits_{i}|m_{i}|$ (i=1,2,3) is found to be consistent with that of the value given by latest Planck cosmological data, $\sum\limits_{i}|m_{i}|<$0.12 eV. The model further predicts effective neutrino masses for neutrinoless double beta decay, 4.15 meV $\leq m_{ee}\leq$ 30.6 meV, tritium beta decay, 8.4 meV $\leq m_{\beta}\leq$ 30.5 meV, Jarlskog invariant, $J_{CP}=\pm 0.022$ for CP violation, baryon asymmetry $Y_{B}=1.15 \times 10^{-10}$ for normal hierarchical case; and also 49.5 meV $\leq m_{ee}\leq$ 51.7 meV, 49.5 meV $\leq m_{\beta}\leq$ 51.4 meV,  $J_{CP}=\pm 0.022$, $Y_{B}=1.12\times 10^{-10}$ for inverted hierarchical case respectively.

$\emph{keywords:}$ {type-I seesaw, right-handed neutrinos, effective neutrino mass, Jarlskog invariant, baryon asymmetry.}
\end{abstract}

\section{Introduction}

Since the observation of non-zero tiny neutrino masses as well as non-zero $\theta_{13}$ neutrino mixing angle in several neutrino oscillation experiments \cite{ref1,ref2,ref3}, many neutrino mass models  based on different discrete symmetries such as $S_{3}$, $S_{4}$, $ A_{4}$, $A_{5}$ etc.\cite{ref4,ref5,ref6,ref7,ref8}, have been developed. In most of the cases, the Standard Model(SM) has been extended upto desired symmetries by adding suitable field contents with respective charges. The current neutrino oscillation experimental data provided by NuFIT \cite{ref9}, is given in Table \ref{t1}. \\

\begin{table}[b]
\begin{center}
\begin{tabular}{c | c }
 
\hline
NH & IH  \\
\hline

$sin^{2} \theta_{12}$ = [0.270, 0.341] & $sin^{2} \theta_{12}$ = [0.270, 0.341] \\

$sin^{2} \theta_{13}$ = [0.02029, 0.2391] & $sin^{2} \theta_{13}$ = [0.02047, 0.02396] \\

$sin^{2} \theta_{23}$ = [0.406, 0.620] & $sin^{2} \theta_{23}$ = [0.412, 0.623] \\

$\delta = [108^{\circ}, 404^{\circ}]$ & $\delta = [192^{\circ}, 360^{\circ}]$ \\

$\Delta m_{21}^{2}$=[6.82, 8.03]$\times 10^{-5}eV^{2}$ & $\Delta m_{21}^{2}$=[6.82, 8.03]$\times 10^{-5}eV^{2}$ \\

$\Delta m_{31}^{2}$=[+2.428, +2.597]$\times 10^{-3}eV^{2}$ & $\Delta m_{32}^{2}$=[-2.581, -2.408]$\times 10^{-3}eV^{2}$ \\
\hline

\end{tabular}
\end{center}
\caption{NuFIT data on neutrino oscillation parameters} 
\end{table}\label{t1}

Among the various discrete symmetries\cite{ref4,ref5,ref6,ref7,ref8}, $\Delta$(27)\cite{ref10,ref11,ref12,ref13,ref14,ref15,ref16,ref17} is one of the possible discrete symmetries so far used to describe the observed pattern of SM lepton masses and mixing angles. Unlike any other discrete symmetries, the $\Delta(27)$ discrete symmetry can lead to complex VEVs with calculable phases stable against radiative corrections and such calculable phases contribute to geometrical spontaneous CP violation \cite{ref18,ref19,ref20}. The observed neutrino mixing pattern is very closed to the tribimaximal (TBM) mixing pattern \cite{ref21,ref22} . However, the only difference is with the value of $\theta_{13}$ mixing angle, and $\Delta$(27) symmetry is a suitable discrete symmetry to generate non-zero $\theta_{13}$ value. The non-zero $\theta_{13}$ value can be obtained by perturbing the neutrino mass matrix upto first order, generating a perturbed mass matrix which in turn contributes to both the eigenvalues and eigenvectors of the neutrino mass matrix. The method of using type-I seesaw mechanism to produce neutrino mass hierarchy and tiny neutrino masses based on $\Delta$(27) discrete symmetry, was proposed in Ref.\cite{ref23} which has some differences with the present work. In Ref.\cite{ref23}, (a) the three right-handed charged leptons are put in $1_{1}$, $1_{2}$, $1_{3}$ representations under $\Delta$(27) symmetry, (b) out of the two $\Delta$(27) scalar triplets, one is put $\bf{2}$ representation under $SU(2)_{L}$ symmetry, (c) the VEVs of the scalars are complex in nature, (d) the charged lepton mass matrix is diagonalized by a specific form of unitary matrix etc.  Furthermore, the way of solving the effective neutrino mass matrix to produce tiny neutrino masses as well as mixing angles in the present work, is very different from the way that is calculated in Ref.\cite{ref23}. In the present work, we can generate the experimental values of neutrino oscillation parameters consistent with the latest Planck cosmological data\cite{ref24}, $\sum\limits_{i}|m_{i}|<$0.12 eV  by  considering minimal number of scalars. \\

In the present work, we use  type-I seesaw mechanism within the framework of $\Delta$(27) discrete symmetry, and study its phenomenological implications to non-zero tiny neutrino mass, non-zero $\theta_{13}$ value, Jarlskog invariant ($J_{CP}$), effective neutrino mass parameters, baryon asymmetry through unflavored thermal leptogenesis, which are consistent with latest Planck cosmological data on the sum of three absolute neutrino mass eigenvalues. \\

The paper is organised in the following way. We give a description of the model and its related Lagrangian in section 2. Numerical analysis is presented in section 3. Section 4 contains a brief analysis of effective neutrino mass parameters for neutrinoless double beta decay, tritium beta decay and Jarlskog invariant for CP violation, followed by unflavoured thermal leptogenesis to obtain baryon asymmetry in section 5. Section 6 deals with summary and conclusion. \\

\section{Description of the Model under $\Delta(27)$}
The Standard Model gauge group is extended to additional discrete flavor symmetry $\Delta$(27) where we introduce two additional Higgs doublets, three right-handed neutrinos and two triplets ($\phi$, $\chi$) under $\Delta$(27), to generate tiny non-zero neutrino masses and non-zero $\theta_{13}$ value. The particle content of the proposed model with corresponding charges under SM gauge group and $\Delta$(27) discrete symmetry, is summarized in the Table \ref{t2} . \\

\begin{table}
\begin{center}
\begin{tabular}{c c c c c c c c c}
\hline
Field &  ($L_{eL}, L_{\mu L}, L_{\tau L}$) &  ($e_{R}, \mu_{R}, \tau _{R}$) &  $H_{1}$ & $H_{2}$ & $H_{3}$ & $\nu_{Rj}$ & $\phi$ & $\chi$ \\
\hline
$SU(2)_{L}$ & 2 & 1 & 2 & 2 & 2 & 1 & 1 & 1 \\

$\Delta$(27) & $\bar{3}$ & $\bar{3}$ & $1_{1}$ & $1_{2}$ & $1_{3}$ & 3 & 3 & 3 \\

$Z_{2}$ & 1 & 1 & 1 &  1 & 1 & -1 & -1 & 1 \\
\hline
\end{tabular}
\end{center}
\caption{\footnotesize{Particle content of the proposed model under SM gauge group and $\Delta$(27) symmetry. Here, $j$=1,2,3.}}

\end{table}{\label{t2}}

With the particle content in Table \ref{t2} and the tensor products of $\Delta$(27) discrete symmetry group in Appendix \ref{secA1}, the Yukawa interaction for charged leptons is given by \\
\begin{equation}
-\mathcal{L}_{yuk}^{l}= h_{k}(\overline{L}_{iL}H_{k}l_{jR}),
\end{equation}

leading to the charged lepton mass matrix,\\

\begin{align*}
M_{l}= diag( & h_{1}v_{1}+h_{2}v_{2}+h_{3}v_{3}, h_{1}v_{1}+\omega^{2}h_{2}v_{2}+ \nonumber \\ &\omega h_{3}v_{3}, h_{1}v_{1}+\omega h_{2}v_{2}+\omega^{2}h_{3}v_{3})
\end{align*}
where $v_{1,2,3}$ are the VEVs of $H_{k}(k=1,2,3)$ and $h_{1}$, $h_{2}$, $h_{3}$ are Yukawa couplings respectively.\\

For the neutrino sector, the Lagrangian for neutral leptons invariant under $\Delta$(27) discrete symmetry is given by \\

\begin{equation}
-\mathcal{L}_{yuk}^{\nu}=\frac{1}{\Lambda}y_{k}(\overline{L}_{iL}\nu_{Rj})_{\bar{3}}\phi H_{k} + \frac{\lambda_{i}}{2}(\overline{\nu}_{Ri}^{c}\nu_{Ri})_{\bar{3}}\chi
\end{equation}

where $i$, $j$ vary from 1 to 3, and $\Lambda$ being the cut-off high energy scale. $Z_{2}$ symmetry is introduced in order to remove the terms coming from the interaction of $\phi$ and $\chi$. With the VEVs $\langle \phi \rangle=(v_{\phi},v_{\phi},v_{\phi_{3}})$ and $\langle \chi \rangle=v_{\chi}(1,0,1)$ \cite{ref23, ref17} of the two triplets $\phi$ and $\chi$, the resulting Dirac and Majorana neutrino mass matrices can be written as, \\
\begin{equation*}
m_{D}=\frac{1}{\Lambda}
\begin{pmatrix}
y_{1}v_{1}v_{\phi} & y_{2}v_{2}v_{\phi_{3}} & y_{2}v_{2}v_{\phi} \\
y_{2}v_{2}v_{\phi_{3}} & y_{1}v_{1}v_{\phi} & y_{2}v_{2}v_{\phi} \\
y_{2}v_{2}v_{\phi} &y_{2}v_{2}v_{\phi} & y_{1}v_{1}v_{\phi_{3}} 
\end{pmatrix}
= 
\begin{pmatrix}
x & z & y \\
z & x & y \\
y & y & t
\end{pmatrix}
\end{equation*}

\begin{equation*}
M_{R}= \frac{1}{2}
\begin{pmatrix}
\lambda_{1}v_{\chi} & \lambda_{2}v_{\chi}  & 0 \\
\lambda_{2}v_{\chi}  & 0 & \lambda_{2}v_{\chi}  \\
0 & \lambda_{2}v_{\chi}  &\lambda_{1}v_{\chi} 
\end{pmatrix}
=
\begin{pmatrix}
a & b & 0 \\
b & 0 & b \\
0 & b & a
\end{pmatrix}
\end{equation*}

where $x= \frac{y_{1}v_{1} v_{\phi}}{\Lambda}$, $y=\frac{y_{2}v_{2}v_{\phi}}{\Lambda}$, $z=\frac{y_{2}v_{2} v_{\phi_{3}}}{\Lambda} $,  $t=\frac{y_{1}v_{1} v_{\phi_{3}}}{\Lambda}$, $a=\frac{\lambda_{1} v_{\chi}}{2}$, $b=\frac{\lambda_{2} v_{\chi}}{2}$. \\

 The effective neutrino mass matrix is generated through the type-I seesaw mechanism,
$M_{\nu}= -m_{D}^{T}M_{R}^{-1}m_{D}$, as \\

\begin{equation}
M_{\nu}= \frac{1}{ab^2}
\begin{pmatrix}
m_{11} & m_{12} & m_{13} \\
m_{21} & m_{22} & m_{23} \\
m_{31} & m_{32} & m_{33} 
\end{pmatrix}
\end{equation}
where \\ $m_{11} = -b^2(x-y)^2-2ab(x+y)z+a^2z^2$, \\
 $m_{12}= b^{2}(x-y)(y-z)+a^{2}xz-ab(x^{2}+xy+yz+z^{2})$, \\
 $m_{13} = b^{2}(t-y)(x-y)+a^{2}yz-ab(y^{2}+xy+tz+yz)$,\\
 $m_{21}=m_{12}$, \\
 $m_{22}= a^{2}x^{2}-b^{2}(y-z)^{2}-2abx(y+z)$, \\
 $m_{23}=a^{2}xy-b^{2}(t-y)(y-z)-ab(tx+xy+y^{2}+yz)$,\\
 $m_{31}=m_{13}$, \\
 $m_{32}=m_{23}$, \\
 $m_{33}=a^{2}y^{2}-b^{2}(t-y)^{2}-2aby(t+y)$.\\

The expression of $M_{\nu}$ can also be written as \\
\begin{align}\label{eq1}
M_{\nu}&=
\begin{pmatrix}
A & B & B \\
B & C & D \\
B & D & C 
\end{pmatrix}
+ 
\begin{pmatrix}
0 & a_{1} & b_{1} \\
a_{1} & a_{2} & 0 \\
b_{1} & 0 & b_{2}
\end{pmatrix} 
\nonumber \\ &\equiv M^{\prime}+\delta M^{\prime}
\end{align}

where \\
\begin{equation*}
A= \frac{-b^2(x-y)^2-2ab(x+y)z+a^2z^2}{2 a b^{2}}, \\
\end{equation*}
\begin{equation*}
B= \frac{-ab(xy+yz)}{2a b^{2}},\hspace{0.5cm} C=\frac{-b^{2}y^{2}}{2a b^{2}}, \\
\end{equation*}
\begin{equation}
D=\frac{a^{2}xy-b^{2}(t-y)(y-z)-ab(tx+xy+y^{2}+yz)}{2a b^{2}}, \\
\end{equation}
\begin{equation*}
a_{1}=\frac{b^{2}(x-y)(y-z)+a^{2}xz-ab(x^{2}+z^{2})}{2a b^{2}}, \\
\end{equation*}
\begin{equation*}
b_{1}=\frac{b^{2}(t-y)(x-y)+a^{2}yz-ab(y^{2}+tz)}{2a b^{2}}, \\
\end{equation*}
\begin{equation*}
a_{2}=\frac{a^{2}x^{2}-2abx(y+z)-b^{2}(z^{2}-2yz)}{2a b^{2}}, \\
\end{equation*}
\begin{equation}
b_{2}=\frac{a^{2}y^{2}-2aby(t+y)-b^{2}(t^{2}-2yt)}{2a b^{2}}. \\
\end{equation}

The first matrix in eq.(\ref{eq1}) is invariant under $\mu - \tau$ symmetry while the second one is a perturbed matrix which can lead to non-zero $\theta_{13}$ value and CP violating phase in lepton sector. The three mass eigenvalues and the corresponding mixing matrix of the first matrix in eq.(\ref{eq1}) are given by \\
\begin{align}
& m_{1,2}^{\prime}=\frac{1}{2}\big[A+C+D\pm\sqrt{(A-C-D)^{2}+8B^{2}} \big] \hspace{0.2cm} , 
\nonumber \\ & m_{3}^{\prime}= C-D
\end{align}
\begin{equation}\label{eq2}
U^{\prime}=
\begin{pmatrix}
cos \theta & sin \theta & 0 \\
-\frac{sin \theta}{\sqrt{2}} & \frac{cos \theta}{\sqrt{2}}  & \frac{-1}{\sqrt{2}} \\
-\frac{sin \theta}{\sqrt{2}} & \frac{cos \theta}{\sqrt{2}}  & \frac{1}{\sqrt{2}} \\
\end{pmatrix}
\end{equation}
where $\theta=\arcsin \big(\frac{\mathcal{K}}{\sqrt{\mathcal{K}^{2}+2}}\big)$ ; $\mathcal{K}=\frac{A-C-D-\sqrt{(A-C-D)^{2}+8B^{2}}}{2B}$. The mixing matrix $U^{\prime}$ implies that $\theta_{13}=0$. However, we can generate non-zero $\theta_{13}$ value by considering $\delta M^{\prime}$
of eq.(\ref{eq1}) as perturbed matrix. By using first order perturbation theory, the eigenvalues of the matrix (\ref{eq1}) are given by \\

\begin{equation}
m_{1}=m_{1}^{\prime}+\frac{1}{2}sin \theta \big[ (a_{2}+b_{2})sin \theta-2\sqrt{2}(a_{1}+b_{1})cos \theta \big] ,
\end{equation}
\begin{equation}
m_{2}=m_{2}^{\prime}+\frac{1}{2} cos \theta \big[ (a_{2}+b_{2})cos \theta+2\sqrt{2}(a_{1}+b_{1})sin \theta \big] \hspace{0.3cm}  
\end{equation}
\begin{equation}
m_{3}=m_{3}^{\prime}+\frac{1}{2}(a_{2}+b_{2}).
\end{equation} 

The corresponding mixing matrix of eq.(\ref{eq1}) is given by \\

\begin{equation}\label{eq3}
U=U^{\prime}+\delta U^{\prime}
\end{equation}
where $U^{\prime}$ is given in (\ref{eq2}) and the elements of $\delta U^{\prime}$ are given in Appendix \ref{secA2}. The neutrino mixing angles $\theta_{12}$, $\theta_{13}$, $\theta_{23}$ and Dirac CP-phase, $\delta$ are related to the elements of neutrino mixing matrix \cite{ref25}as \\
\begin{align}
& tan \theta_{12}=\frac{|U_{12}|}{|U_{11}|} \hspace{0.3cm};
\hspace{0.3cm} tan \theta_{23}=\frac{|U_{23}|}{|U_{33}|} \hspace{0.3cm}; 
\nonumber \\ & sin \theta_{13}=U_{13}e^{i\delta}.
\end{align}

The mixing elements $U_{11}$ and $U_{12}$ can be written in terms of $\theta$ and $t_{12}$(=tan $\theta_{12}$) using eqs.(\ref{eq2}), (\ref{eq3}) and (\ref{eq4}), \\

\begin{equation}
U_{11}= \frac{1}{cos \theta + t_{12} sin \theta} \hspace{0.3cm}; 
\hspace{0.3cm} U_{12}= \frac{t_{12}}{cos \theta + t_{12} sin \theta}
\end{equation}

If we take $U_{11}=0.812$ as $U_{11}$ $\in$ (0.803,0.845) at 3$\sigma$ confidence level, then it is found that $t_{12}$ = 0.71885, $U_{12}$ = 0.58371 and sin $\theta_{12}$ = 0.58369. These are consistent with the 3$\sigma$ confidence level of neutrino oscillation data. \\

Further, using the relations \\
\begin{equation}
U_{11}=U_{11}^{\prime}+\delta U_{11}^{\prime} \hspace{0.3cm},
\end{equation}

and \begin{equation}
U_{13}=U_{13}^{\prime}+\delta U_{13}^{\prime}   \hspace{0.3cm},
\end{equation}

the terms $b_{1}$ and $a_{1}$ are given by \\

\begin{align}
b_{1}=&\frac{(sin \theta - t_{12}cos \theta)(m_{1}^{\prime}-m_{2}^{\prime})}{\sqrt{2}(cos \theta + t_{12}sin \theta)(cos^{2}\theta-sin^{2}\theta)}+\frac{(a_{2}+b_{2})sin \theta cos \theta}{2\sqrt{2}(cos^{2}\theta - sin^{2}\theta)} + \nonumber \\
&\frac{2\sqrt{2}(m_{3}^{\prime}-m_{1}^{\prime})(m_{3}^{\prime}-m_{2}^{\prime})sin \theta_{13}e^{-i\delta}-(a_{2}-b_{2})(m_{1}^{\prime}-m_{2}^{\prime})sin \theta cos \theta}{4(-m_{1}^{\prime}sin^{2}\theta - m_{2}^{\prime}cos^{2}\theta + m_{3}^{\prime})},
\end{align}
\begin{align}
a_{1}=b_{1}-\frac{2\sqrt{2}(m_{3}^{\prime}-m_{1}^{\prime})(m_{3}^{\prime}-m_{2}^{\prime})sin \theta_{13}e^{-i\delta}-(a_{2}-b_{2})(m_{1}^{\prime}-m_{2}^{\prime})sin \theta cos \theta}{2(-m_{1}^{\prime}sin^{2}\theta - m_{2}^{\prime}cos^{2}\theta + m_{3}^{\prime})}.
\end{align}

\section{Numerical Analysis}
For detailed numerical analysis, we choose a random range on the model parameters $a_{2}$ and $b_{2}$ for both normal and inverted hierarchy to study the correlation among the model parameters consistent with the 3$\sigma$ confidence level of current neutrino oscillation data and latest Planck cosmological data on the sum of the three absolute neutrino masses. 

\subsection{Normal hierarchy}\label{subsec2}
\begin{figure}[htbp]
\begin{subfigure}{0.5\textwidth}
  \centering
    \includegraphics[scale=0.5]{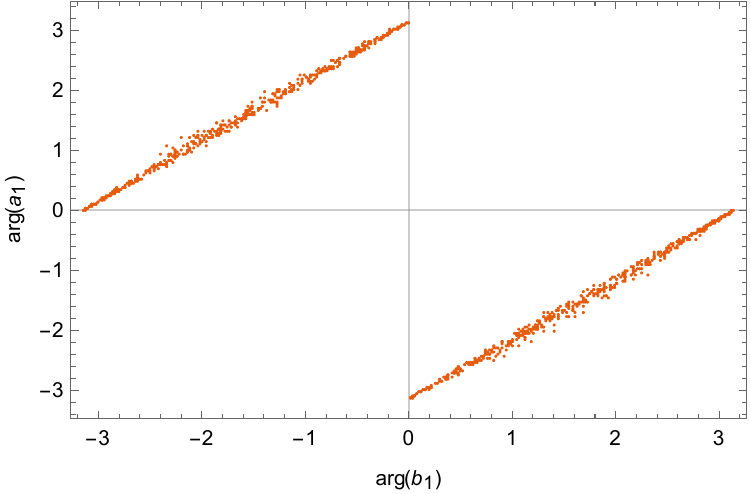}
    \caption{}
    \end{subfigure}
    \begin{subfigure}{0.5\textwidth}
 \includegraphics[scale=0.5]{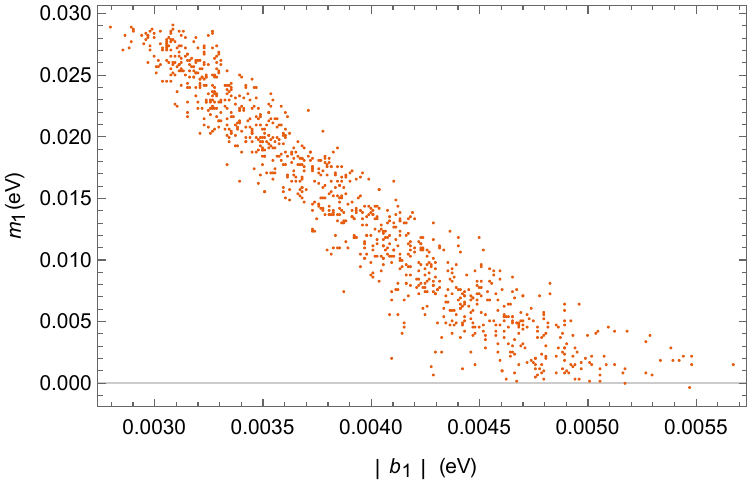}
    \caption{}
    \end{subfigure}
 \caption{\footnotesize{Plot representing the variation between the model parameters (a) $arg(a_{1})$ and $arg(b_{1})$, (b) $|m_{1}|$ and $|b_{1}|$ for normal hierarchy.}}\label{f1}
\end{figure}

\begin{figure}[htbp]
\begin{subfigure}{0.5\textwidth}
  \centering
    \includegraphics[scale=0.5]{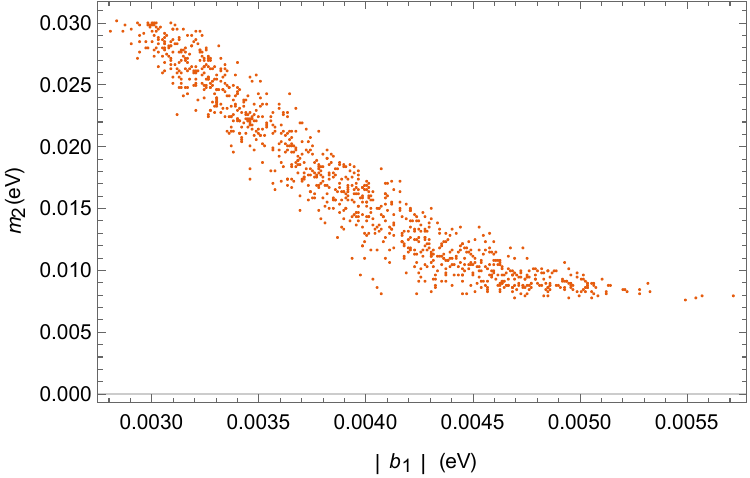}
    \caption{}
    \end{subfigure}
    \begin{subfigure}{0.5\textwidth}
 \includegraphics[scale=0.5]{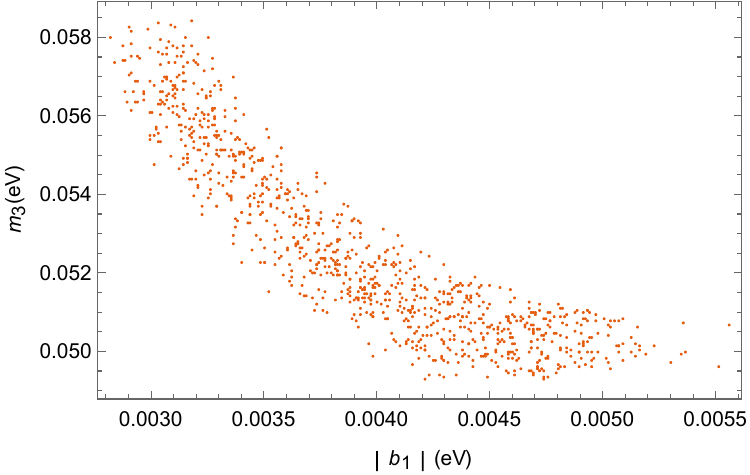}
    \caption{}
    \end{subfigure}
 \caption{\footnotesize{Plot representing the variation between the model parameters (a) $m_{2}$ and $|b_{1}|$, (b) $m_{3}$ and $|b_{1}|$ for normal hierarchy.}}\label{f2}
\end{figure}

\begin{figure}[t]
\centering
\includegraphics[scale=0.5]{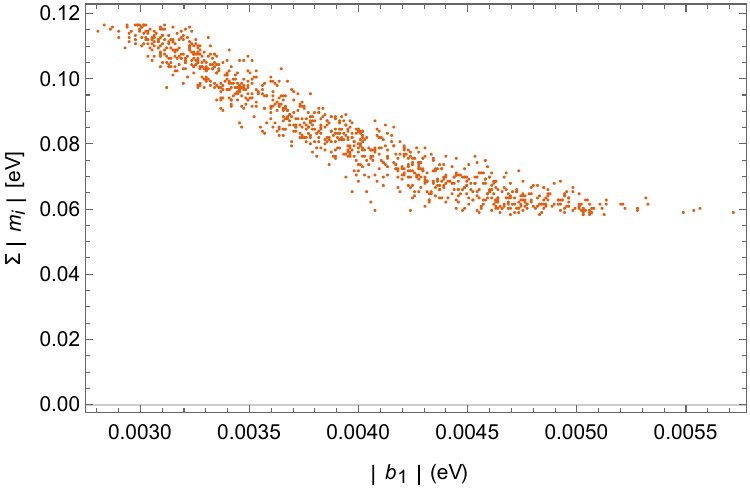}
\caption{\footnotesize{Plot representing the variation between the model parameters $\sum |m_{i}|$ and $|b_{1}|$ for normal hierarchical case.}}\label{f3}
\end{figure}

In case of normal hierarchy (NH), we choose $a_{2}\in$ [-1$\times 10^{-5}$, -9$\times 10^{-5}$] eV, $b_{2}\in$ [1$\times 10^{-5}$, 9$\times 10^{-5}$] eV, $m_{1}^{\prime}\in$ [0.0001, 0.029] eV. Here, the upper bound on $m_{1}^{\prime}$ is due to the constraint on the sum of three absolute neutrino masses given by latest Planck cosmological data. We also choose sin $\theta=\frac{1}{\sqrt{3}}$ and cos $\theta = \sqrt{\frac{2}{3}}$ as $U^{\prime}$ of eq.($\ref{eq2}$) is very closed to TBM mixing matrix. With $\delta \in [108^{\circ},404^{\circ}]$ and $s_{13}\in $ [0.142, 0.155], we plot the correlation among the model parameters. Fig. $\ref{f1}(a)$ represents the variation of arg($a_{1}$) and arg($b_{1}$) showing that these parameters are opposite in phase. Figs. $\ref{f1}(b)$ and $\ref{f2}$ represent the variations of $m_{1}$, $m_{2}$ and $m_{3}$ with $|b_{1}|$ showing the allowed range of $m_{1}$, $m_{2}$ and $m_{3}$. We plot the variation of $\sum|m_{i}|$ with $|b_{1}|$ in Fig. $\ref{f3}$. Here, it is found that the output value of $\sum|m_{i}|$ is consistent with that of the value given by Planck cosmological data.  \\

Table \ref{t3} represents the values of other model parameters for the input values $a_{2}= -5.2\times 10^{-5}$eV, $b_{2}=6.9 \times 10^{-5}$ eV, $m_{1}^{\prime}$=0.01eV, $\delta$=2.74 rad($157^{\circ}$) for normal hierarchical case. In our model, we choose random parameters $a_{2}<$ $ 10^{-4}$ eV and $b_{2}<$ $ 10^{-4}$ eV so as to get experimentally allowed neutrino oscillation parameters. The magnitude of the leptonic mixing matrix is found to be \\

\begin{equation}
|U|=
\begin{pmatrix}
0.812 & 0.58371 & 0.1485 \\
0.3302 & 0.6391 & 0.7018\\
0.4949 & 0.5096 & 0.7124
\end{pmatrix}
\end{equation}
leading to tan $\theta_{12}$=0.71885, tan $\theta_{23}$=0.98511.

\begin{table}
\begin{center}
\begin{tabular}{c c}
\hline
Parameters & Output values \\
\hline
\hline
$a_{1}$ & (-4.19 $\times 10^{-3}$ - 0.00034$\emph{i}$) eV \\
$b_{1}$ & (4.3 $\times 10^{-3}$ + 0.00034$\emph{i}$) eV \\
$m_{1}$ & 0.00992 eV \\
$m_{2}$ & 0.01332 eV \\
$m_{3}$ & 0.05182 eV \\
$\Delta m_{21}^{2}$ & 7.89 $\times 10^{-5} eV^{2}$ \\
$\Delta m_{31}^{2}$ & 2.51 $\times 10^{-3} eV^{2}$ \\
$\sum\limits_{i} |m_{i}|$ & 0.0751 eV \\
$m_{ee}$ & 12.2 meV \\
$m_{\beta}$ & 13.6 meV \\
\hline

\end{tabular}
\end{center}
\caption{\label{t3}\footnotesize{Model parameter output values for $a_{2}=-5.2\times 10^{-5}$ eV, $b_{2}=6.9 \times 10^{-5}$ eV, $m_{1}^{\prime}$=0.01 eV and $\delta=2.74$ rad ($157^{\circ}$) for normal hierarchy.}}
\end{table}

\newpage
\subsection{Inverted hierarchy}
\begin{figure}[htbp]
\begin{subfigure}{0.5\textwidth}
  \centering
    \includegraphics[scale=0.5]{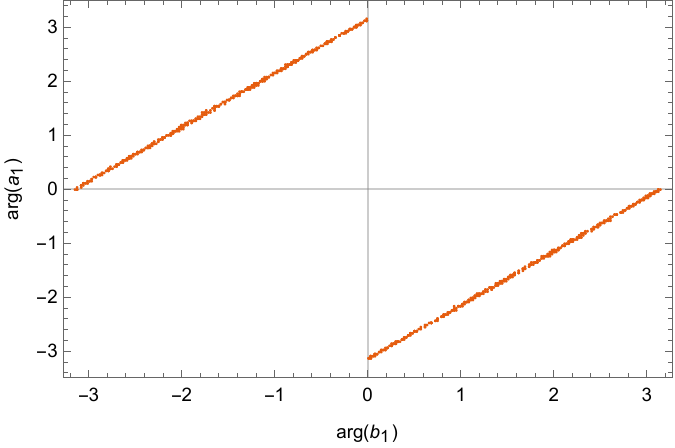}
    \caption{}
    \end{subfigure}
    \begin{subfigure}{0.5\textwidth}
 \includegraphics[scale=0.5]{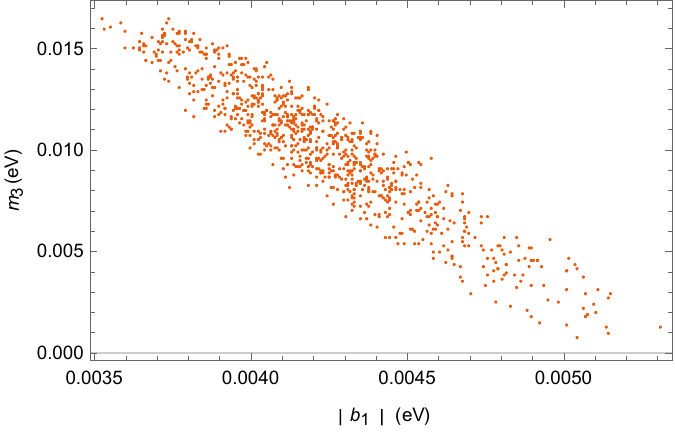}
    \caption{}
    \end{subfigure}
 \caption{\footnotesize{Plot representing the variation between the model parameters (a) $arg(a_{1})$ and $arg(b_{1})$, (b) $|m_{3}|$ and $|b_{1}|$ for inverted hierarchy.}}\label{f4}
\end{figure}

\begin{figure}
\centering
\includegraphics[scale=0.5]{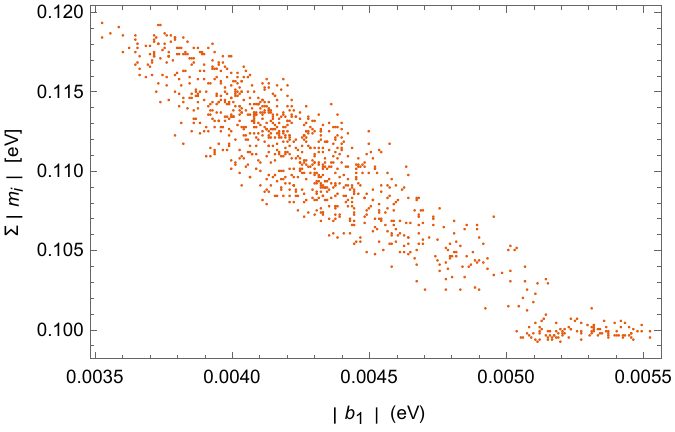}
\caption{\footnotesize{Plot representing the variation between the model parameters $\sum |m_{i}|$ and $|b_{1}|$ for inverted hierarchical case.}}\label{f5}
\end{figure}

In case of inverted hierarchy (IH), we choose $a_{2}\in$ [-1$\times 10^{-5}$, $10^{-4}$] eV, $b_{2}\in$ [1$\times 10^{-5}$, 9$\times 10^{-5}$] eV,  $m_{2}^{\prime}\in$ [0.05, 0.0518] eV. With $\delta \in$ [$192^{\circ}, 360^{\circ}$], we plot the correlations among the model parameters. Fig. $\ref{f4}(a)$ represents the variation of arg($a_{1}$) and arg($b_{1}$). In inverted hierarchy also, it is found that these parameters are opposite in phase. Figs. $\ref{f4}(b)$ represents the variation of $m_{3}$ with $|b_{1}|$ showing the allowed range of  $m_{3}$.  We plot the variation of $\sum|m_{i}|$ with $|b_{1}|$ in Fig. $\ref{f5}$. Here, it is found that the output value of $\sum|m_{i}|$ is consistent with that of the value given by Planck cosmological data.  \\

Table \ref{t4} represents the values of other model parameters for the input values $a_{2}= -5.2\times 10^{-5}$eV, $b_{2}=6.9 \times 10^{-5}$ eV, $m_{2}^{\prime}$ = 0.05eV, $\delta$ = 2.74 rad($157^{\circ}$) for inverted hierarchical case. With this, the magnitude of the leptonic mixing matrix is found to be \\

\begin{equation}
|U|=
\begin{pmatrix}
0.812 & 0.58371 & 0.1485 \\
0.3258 & 0.6332 & 0.7091 \\
0.4991 & 0.5156 & 0.7051
\end{pmatrix}
\end{equation}
leading to tan $\theta_{12}$=0.71885, tan $\theta_{23}$=1.0057.

\begin{table}
\begin{center}
\begin{tabular}{c c}
\hline
Parameters & Output values \\
\hline
\hline
$b_{2}$ & 7.83$\times 10^{-5}$ eV \\
$b_{1}$ & (3.89$\times 10^{-3}$-0.00451$\emph{i}$) eV \\
$a_{1}$ & (-3.89$\times 10^{-3}$+0.00451$\emph{i}$) eV \\
$m_{1}$ & 0.0491993 eV \\
$m_{2}$ & 0.0499912 eV \\
$m_{3}$ & 0.0093142 eV \\
$\Delta m_{21}^{2}$ & 7.85 $\times 10^{-5} eV^{2}$ \\
$\Delta m_{31}^{2}$ & -2.412 $\times 10^{-3} eV^{2}$ \\
$\sum\limits_{i} |m_{i}|$ & 0.1085 eV \\
\hline

\end{tabular}
\end{center}
\caption{\label{t4}\footnotesize{Model parameter output values for $a_{2}=-5.2\times 10^{-5}$ eV, $m_{2}^{\prime}$=0.05 eV and $\delta=2.74$ rad ($157^{\circ}$) for inverted hierarchy.}}
\end{table}

\newpage
\section{Effective neutrino mass parameters and Jarlskog invariant}

The effective neutrino masses for neutrinoless double beta decay \cite{ref26}and tritium beta decay \cite{ref27}are given by \\
\begin{equation*}
m_{ee}=|\sum\limits_{i=1}^{3}U_{ei}^{2}m_{i}| , \\
\end{equation*}
\begin{equation*}
m_{\beta}=\big(\sum\limits_{i=1}^{3}|U_{ei}^{2}|m_{i}^{2}\big)^{1/2}.
\end{equation*}

The Jarlskog invariant is given by \cite{ref28} \\
\begin{align*}
J_{CP}&=Im(U_{e1}U_{\mu 2}U_{e2}^{*}U_{\mu 1}^{*})
\nonumber \\ & =s_{12}c_{12}s_{23}c_{23}c_{13}^{2}s_{13}sin \delta.
\end{align*}

\begin{figure}[htbp]
\begin{subfigure}{0.5\textwidth}
  \centering
    \includegraphics[scale=0.5]{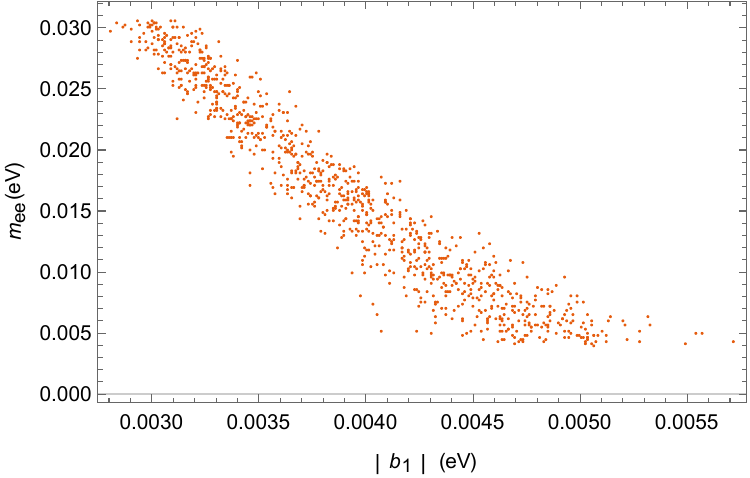}
    \caption{}
    \end{subfigure}
    \begin{subfigure}{0.5\textwidth}
 \includegraphics[scale=0.5]{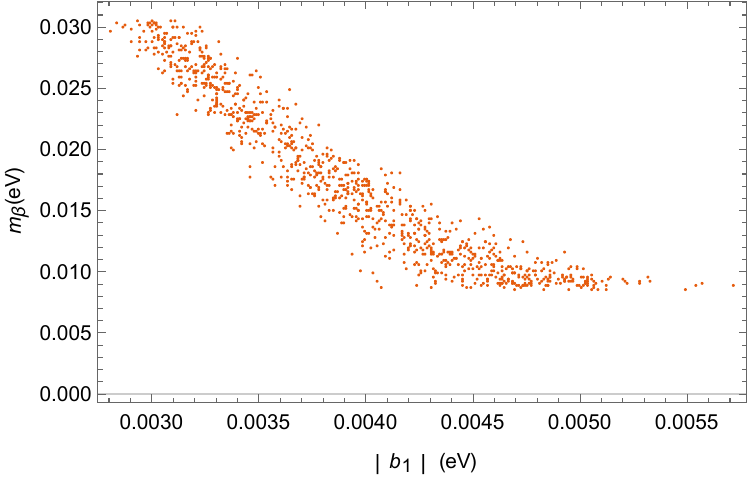}
    \caption{}
    \end{subfigure}
 \caption{\footnotesize{Plot representing the variation between the model parameters (a) $m_{ee}$ and $|b_{1}|$, (b) $m_{\beta}$ and $|b_{1}|$ for normal hierarchy.}}\label{f6}
\end{figure}

\begin{figure}[htbp]
\begin{subfigure}{0.5\textwidth}
  \centering
    \includegraphics[scale=0.5]{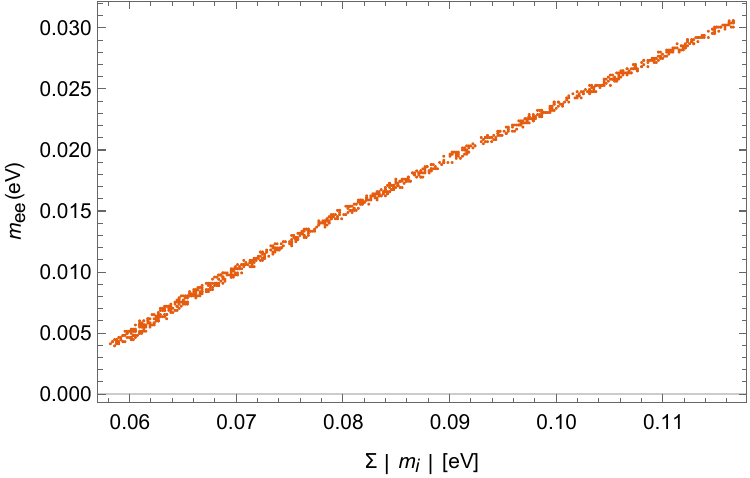}
    \caption{}
    \end{subfigure}
    \begin{subfigure}{0.5\textwidth}
 \includegraphics[scale=0.5]{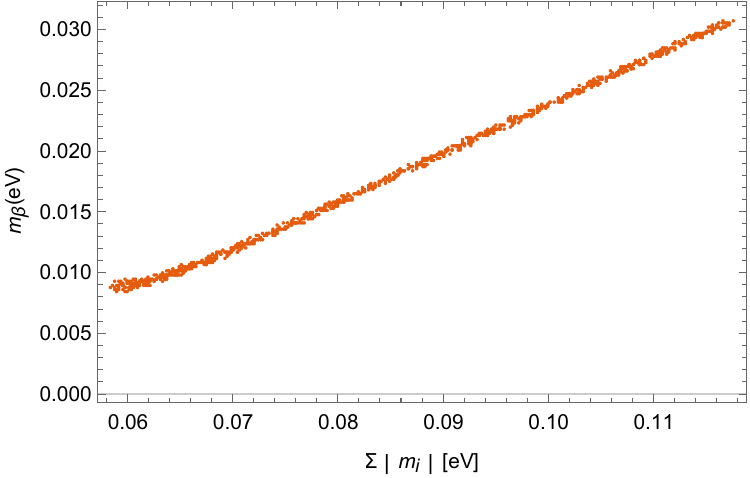}
    \caption{}
    \end{subfigure}
 \caption{\footnotesize{Plot representing the variation between the model parameters (a) $m_{ee}$ and $\sum |m_{i}|$, (b) $m_{\beta}$ and $\sum |m_{i}|$ for normal hierarchy.}}\label{f7}
\end{figure}

\begin{figure}[htbp]
\begin{subfigure}{0.5\textwidth}
  \centering
    \includegraphics[scale=0.5]{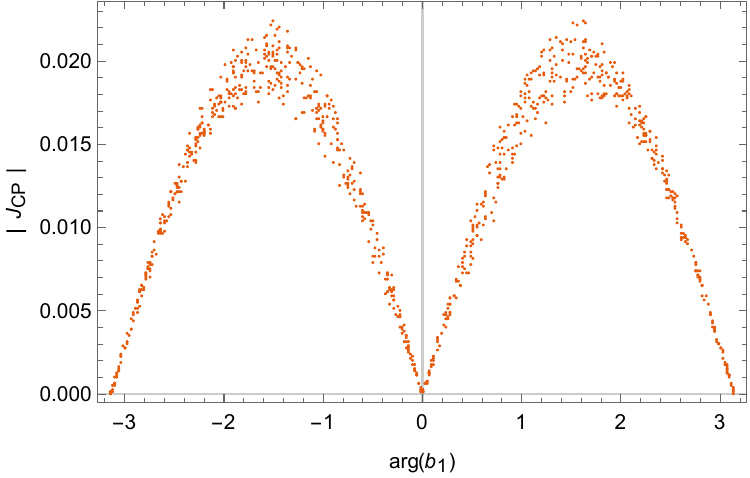}
    \caption{}
    \end{subfigure}
    \begin{subfigure}{0.5\textwidth}
 \includegraphics[scale=0.5]{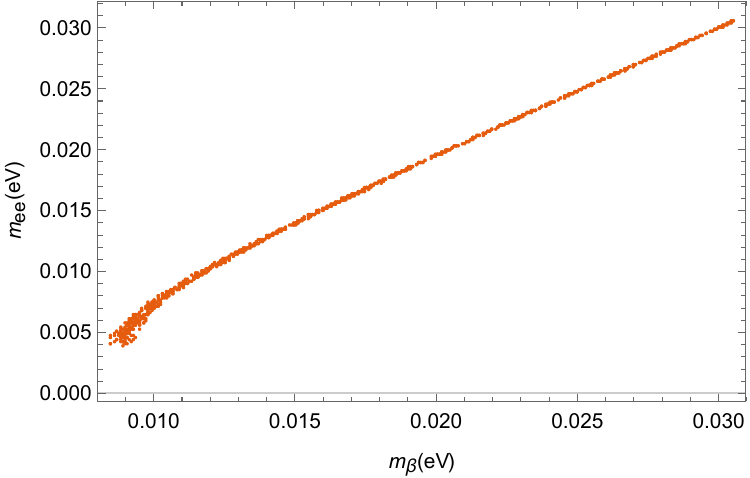}
    \caption{}
    \end{subfigure}
 \caption{\footnotesize{Plot representing the variation between the model parameters (a) $|J_{CP}|$ and $arg(b_{1})$, (b) $m_{ee}$ and $m_{\beta}$  for normal hierarchy.}}\label{f8}
\end{figure}

For NH, Figs. $\ref{f6}, \ref{f7}$ show the dependencies of $m_{ee}$ and $m_{\beta}$ on the parameters $|b_{1}|$ and $\sum |m_{i}|$ . It is found that the theoretically predicted values are  4.15 meV $\leq m_{ee}\leq 30.6$ meV, 8.4 meV $\leq m_{\beta} \leq 30.5$ meV while the Jarlskog invariant is found to be -0.0219 $\leq J_{CP} \leq 0.0221$ which is plotted in Fig. $\ref{f8}(a)$. We also plot the variation of $m_{ee}$ and $m_{\beta}$ in Fig. $\ref{f8}(b)$.

\begin{figure}[htbp]
\begin{subfigure}{0.5\textwidth}
  \centering
    \includegraphics[scale=0.5]{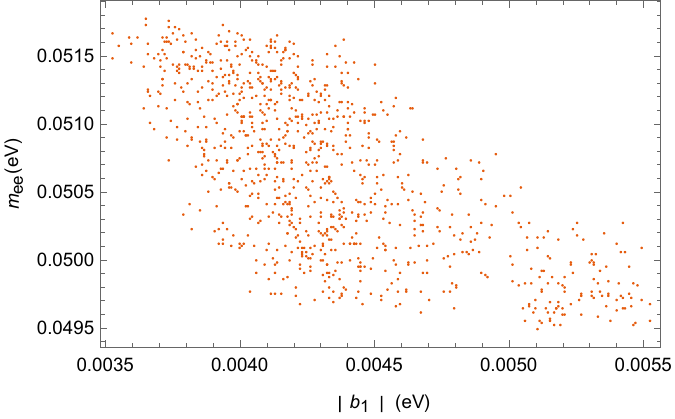}
    \caption{}
    \end{subfigure}
    \begin{subfigure}{0.5\textwidth}
 \includegraphics[scale=0.5]{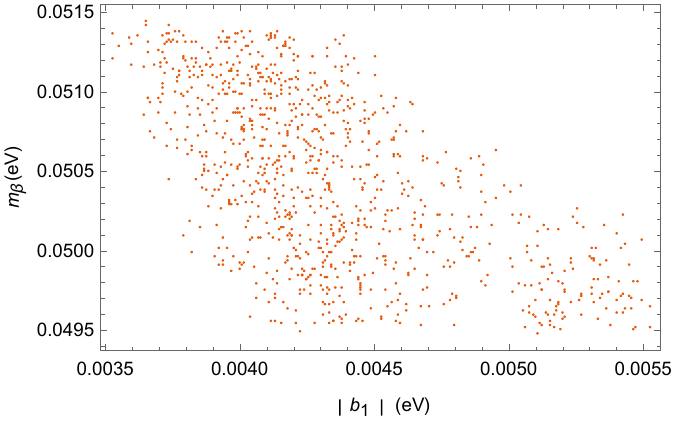}
    \caption{}
    \end{subfigure}
 \caption{\footnotesize{Plot representing the variation between the model parameters (a) $m_{ee}$ and $|b_{1}|$, (b) $m_{\beta}$ and $|b_{1}|$ for inverted hierarchy.}}\label{f9}
\end{figure}

\begin{figure}[htbp]
\begin{subfigure}{0.5\textwidth}
  \centering
    \includegraphics[scale=0.5]{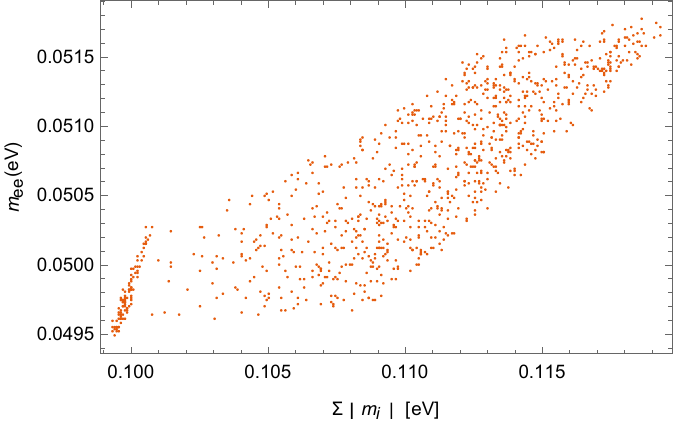}
    \caption{}
    \end{subfigure}
    \begin{subfigure}{0.5\textwidth}
 \includegraphics[scale=0.5]{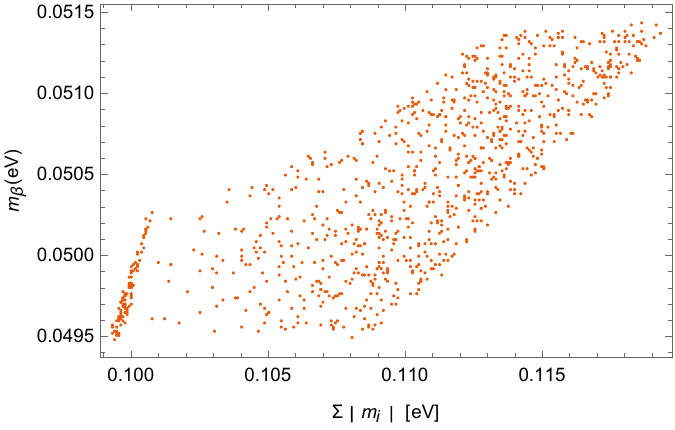}
    \caption{}
    \end{subfigure}
 \caption{\footnotesize{Plot representing the variation between the model parameters (a) $m_{ee}$ and $\sum |m_{i}|$, (b) $m_{\beta}$ and $\sum |m_{i}|$ for inverted hierarchy.}}\label{f10}
\end{figure}

\begin{figure}[htbp]
\begin{subfigure}{0.5\textwidth}
  \centering
    \includegraphics[scale=0.5]{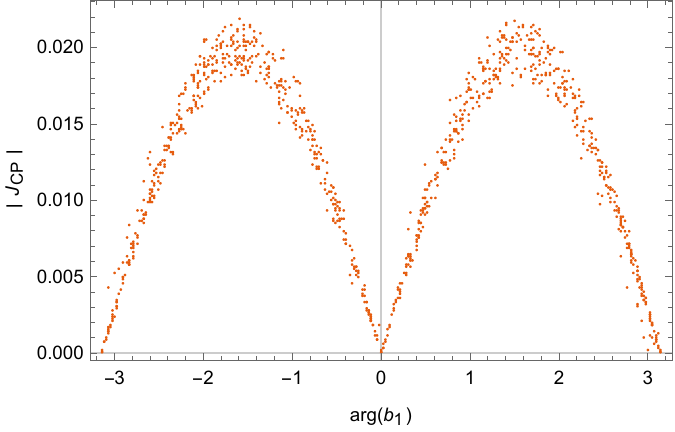}
    \caption{}
    \end{subfigure}
    \begin{subfigure}{0.5\textwidth}
 \includegraphics[scale=0.5]{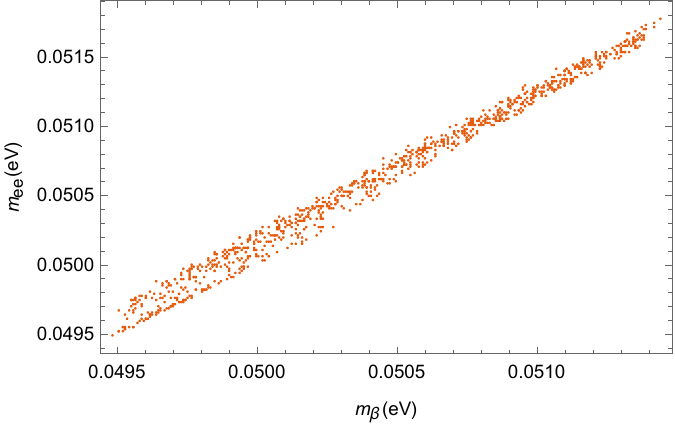}
    \caption{}
    \end{subfigure}
 \caption{\footnotesize{Plot representing the variation between the model parameters (a) $|J_{CP}|$ and $arg(b_{1})$, (b) $m_{ee}$ and $m_{\beta}$  for inverted hierarchy.}}\label{f11}
\end{figure}

For IH, Figs. $\ref{f9}, \ref{f10}$ show the dependencies of $m_{ee}$ and $m_{\beta}$ on the parameters $|b_{1}|$ and $\sum |m_{i}|$. It is found that the theoretically predicted values are  49.5 meV $\leq m_{ee}\leq 51.7$ meV, 49.5 meV $\leq m_{\beta} \leq 51.4$ meV while the Jarlskog invariant is found to be -0.0219 $\leq J_{CP} \leq 0.0223$ which is plotted in Fig. $\ref{f11}(a)$. We also plot the variation of $m_{ee}$ and $m_{\beta}$ in Fig. $\ref{f11}(b)$.

\section{Unflavored thermal leptogenesis}

The CP violation in the right-handed Majorana neutrino decays can significantly contribute to the lepton asymmetry as explained in Ref. \cite{ref29,ref30}. Taking into account the effects of the tree level and one-loop level, the CP violating parameter takes the form \\
\begin{equation}
\epsilon_{i}= \frac{1}{8\pi(h^{\dagger}h)_{ii}}\sum\limits_{j\neq i}Im\big[(h^{\dagger}h)_{ij}\big]^{2}\big[f\big(\frac{M_{j}^{2}}{M_{i}^{2}}\big) + g\big(\frac{M_{j}^{2}}{M_{i}^{2}}\big)\big]
\end{equation}
where the function $f(x)$ and $g(x)$ with $x=\frac{M_{j}^{2}}{M_{i}^{2}}$ come from one-loop vertex and self-energy contributions respectively.
 Considering the CP violating decay of the lightest Majorana neutrino $M_{1}$ and for $x\gg$ 1, the CP violating parameter can be written as \cite{ref31,ref32}  \\
\begin{equation}
\epsilon=-\frac{3}{16\pi}\big[\frac{Im[(h^{\dagger}h)_{12}^{2}]M_{1}}{(h^{\dagger}h)_{11}M_{2}}+\frac{Im[(h^{\dagger}h)_{13}^{2}]M_{1}}{(h^{\dagger}h)_{11}M_{3}}\big]
\end{equation}

where $\emph{h}$ is the Yukawa coupling of the Dirac neutrino mass matrix, $m_{D}$ and $M_{1}$, $M_{2}$, $M_{3}$ are the diagonal elements of the right-handed Majorana neutrino mass matrix, $M_{R}$. \\

The CP violating parameter $\epsilon$ is related to the lepton asymmetry parameter, $Y_{L}$ as \\
\begin{equation}
Y_{L}=\big|\frac{\epsilon k_{1}}{g_{*1}}\big|
\end{equation}

where $g_{*1}$ = 108.5 is the effective number of degrees of freedom at temperature T= $M_{1}$, $k_{1}$ is the dilution factor and is given by \cite{ref29}  \\

\hspace{2cm} $k_{1}= - \frac{1}{2\sqrt{K^{2}+9}}$ for $0\leq K \leq 10$ \\
 
with $K= \frac{\tilde{m}}{m^{*}}$ , $\tilde{m}=\frac{(h^{\dagger}h)_{11}v^{2}}{M_{1}}$ and $m^{*}$ is the equilibrium neutrino mass. \\

The baryon asymmetry is related to lepton asymmetry by \cite{ref29,ref33} \\
\begin{equation}
Y_{B}=bY_{L}
\end{equation} 

with $b=\frac{8N_{f}+4N_{H}}{22N_{f}+13N_{H}}$ where $N_{f}$ is the number of fermion generation and $N_{H}$ is the number of Higgs doublets. \\

\subsection{Leptogenesis from our model}

To calculate lepton and baryon asymmetry, we choose a basis $U_{R}$ such that $M_{R}^{diag}$ = $U_{R}^{T}M_{R}U_{R}$ = diag($M_{1}$, $M_{2}$, $M_{3}$) \cite{ref34}. Accordingly, $m_{D}$ is transformed to the $U_{R}$ basis by $m_{D}$ $\rightarrow$ $m_{D}^{\prime}$ = $m_{D}U_{R}$ with the Dirac neutrino Yukawa coupling h=$\frac{m_{D}U_{R}}{v}$ where $v$ = 174 GeV. The Majorana neutrino mass matrix, $M_{R}$ then becomes \\

\begin{equation}
M_{R}^{diag}=
\begin{pmatrix}
M_{1} & 0 & 0 \\
0 & M_{2} & 0 \\
0 & 0 & M_{3}
\end{pmatrix}
\end{equation}
\\ 
where $M_{1}$ = a , $M_{2}$ = $\frac{1}{2}(a - \sqrt{a^{2}+8b^{2}})$ and 
$M_{3}$ = $\frac{1}{2}(a + \sqrt{a^{2}+8b^{2}})$. For NH, we choose a=$10^{13}$ GeV, b = $10^{14}$ GeV and obtain $M_{1}$ = $10^{13}$ GeV, $|M_{2}|$ = 1.36$\times 10^{14}$ GeV, $M_{3}$ = 1.46$\times 10^{14}$ GeV. We also use the model parameters from Table $\ref{t3}$ and it is found that \\

\begin{equation}
m_{D}^{\prime}=
\begin{pmatrix}
-0.0098+0.25 i & -0.993+25.23 i & -1.204+30.68 i \\
-0.697+17.7 i & 0.183-4.63 i & -1.385+35.26 i \\
-0.537+13.7 i & -0.664+16.93 i & -2.283+58.11 i
\end{pmatrix}.
\end{equation}

Applying h = $\frac{m_{D}^{\prime}}{v}$, we get $(h^{\dagger}h)_{11}$ = 0.00208, $Im [(h^{\dagger}h)_{12}^{2}]$ = 5.5 $\times 10^{-9}$, $Im [(h^{\dagger}h)_{13}^{2}]$ = -1.17 $\times 10^{-7}$. In the analysis, it is found that $\epsilon$ = 2.19 $\times 10^{-7}$. The baryon asymmetry is obtained as $Y_{B}$ = 1.15 $\times 10^{-10}$ which is consistent with the experimental bound \cite{ref24}. \\

For IH, we choose same values of the parameters a, b as in normal hierarchical case. We use the model parameters from Table $\ref{t4}$ and it is found that \\ 

\begin{equation}
m_{D}^{\prime}=
\begin{pmatrix}
-0.315+7.52 i & 0.245-5.79 i & 0.948-24.13 i \\
-0.072+1.97 i & -0.171+3.71 i & 1.01-25.59 i \\
-0.236+6.21 i & -0.064+1.74 i & 0.481-12.31 i
\end{pmatrix}.
\end{equation}

In inverted hierarchical case, we get $(h^{\dagger}h)_{11}$ = 0.00328, $Im [(h^{\dagger}h)_{12}^{2}]$ = -2.13 $\times 10^{-9}$, $Im [(h^{\dagger}h)_{13}^{2}]$ = -1.69 $\times 10^{-7}$ and $\epsilon$ = 2.13 $\times 10^{-7}$. The baryon asymmetry is found to be $Y_{B}$ = 1.12 $\times 10^{-10}$ which is also consistent with the experimental bound.\\

\section{Summary and Conclusion}\label{sec4}
We have discussed neutrino phenomenology by studying type-I seesaw mechanism with $\Delta$(27) symmetry group. The SM gauge group has been extended by adding additional two Higgs doublets, three right-handed neutrinos and two scalar triplets. The Higgs doublets account for charged- lepton mass and along with one scalar triplet, they account for Dirac neutrino mass. The other scalar triplet leads to heavy Majorana neutrino mass. With these, we construct a neutrino mass matrix consisting of a matrix which is invariant under $\mu$ - $\tau$ symmetry and a perturbed matrix.\\

 Further, we succeed to obtain neutrino mixing matrix consistent with recent experimental findings for both normal and inverted hierarchy. In addition, the model also helps in finding effective neutrino mass parameters, $m_{ee}$, $m_{\beta}$, Jarlskog invariant, $J_{CP}$ for CP violation as well as in calculating baryon asymmetry, $Y_{B}$ through unflavoured thermal leptogenesis for both the hierarchical case. Also the baryon asymmetry $Y_{B}$ is consistent to the current experimental bound in both the hierarchical case. In this sense, the analysis show favourable to both the hierarchical case. \\

\section*{Acknowledgements}
Ph. Wilina would like to thank Manipur University for granting Fellowship for Ph. D. program.

\section*{Appendix}
\numberwithin{equation}{section}
\numberwithin{table}{section}
\begin{appendices}
\section{Tensor product rules for $\Delta(27)$ symmetry}\label{secA1}
The $\Delta(27)$ discrete symmetry consists of 9 one-dimensional representation, $1_{i} (i=1,...,9)$ and 2 three-dimensional irreducible representations 3 and $\bar{3}$. Let $(x_{1}, x_{2}, x_{3})$ and $(y_{1}, y_{2}, y_{3})$ be the triplets of $\Delta(27)$ symmetry. Then the tensor products of these triplets are \cite{ref14,ref35} \\

\begin{align*}
\begin{pmatrix}
x_{1}\\
x_{2}\\
x_{3} 
\end{pmatrix}_{3}\otimes
\begin{pmatrix}
y_{1}\\
y_{2}\\
y_{3}
\end{pmatrix}_{3}= 
\begin{pmatrix}
x_{1}y_{1}\\
x_{2}y_{2}\\
x_{3}y_{3}
\end{pmatrix}_{\bar{3}}\oplus
\begin{pmatrix}
x_{2}y_{3}+x_{3}y_{2}\\
x_{3}y_{1}+x_{1}y_{3}\\
x_{1}y_{2}+x_{2}y_{1}
\end{pmatrix}_{\bar{3}} \oplus
\begin{pmatrix}
x_{2}y_{3}-x_{3}y_{2}\\
x_{3}y_{1}-x_{1}y_{3}\\
x_{1}y_{2}-x_{2}y_{1}
\end{pmatrix}_{\bar{3}}
\end{align*}

and \begin{equation}
\begin{pmatrix}
x_{1}\\
x_{2}\\
x_{3}
\end{pmatrix}_{3}\otimes
\begin{pmatrix}
y_{1}\\
y_{2}\\
y_{3}
\end{pmatrix}_{\bar{3}}=
\sum\limits_{i=1}^{9} 1_{i}
\end{equation}

where \\

  $1_{1}=x_{1}\bar{y}_{1}$ + $x_{2}\bar{y}_{2}$ + $x_{3}\bar{y}_{3}$ ; \hspace{0.4cm} $1_{2}=x_{1}\bar{y}$ + $\omega x_{2}\bar{y}_{2}$ + $\omega^{2}x_{3}\bar{y}_{3}$ ; \\
  
$1_{3}=x_{1}\bar{y}$ + $\omega^{2} x_{2}\bar{y}_{2}$ + $\omega x_{3}\bar{y}_{3}$ ; \hspace{0.4cm} $1_{4}= x_{1}\bar{y}_{2}$ + $x_{2}\bar{y}_{3}$ + $x_{3}\bar{y}_{1}$; \\

$1_{5}= x_{1}\bar{y}_{2}$ + $\omega x_{2}\bar{y}_{3}$ + $\omega^{2}x_{3}\bar{y}_{1}$; \hspace{0.4cm} $1_{6}= x_{1}\bar{y}_{2}$ + $\omega^{2} x_{2}\bar{y}_{3}$ + $\omega x_{3}\bar{y}_{1}$; \\

$1_{7}=x_{2}\bar{y}_{1}$ + $x_{3}\bar{y}_{2}$ + $x_{1}\bar{y}_{3}$ ; \hspace{0.4cm} $1_{8}=x_{2}\bar{y}_{1}$ + $\omega^{2} x_{3}\bar{y}_{2}$ + $\omega x_{1}\bar{y}_{3}$ ; \\

$1_{9}=x_{2}\bar{y}_{1}$ + $\omega  x_{3}\bar{y}_{2}$ + $\omega^{2} x_{1}\bar{y}_{3}$; \\

with $\omega = e^{2\pi i/3}$ and 1+ $\omega$ + $\omega^{2}$=0, $\omega^{3}=1$.\\

The singlet multiplications are given in the following table . 

\begin{table}[htbp]
\begin{center}
\begin{tabular}{c| c c c c c c c c}
\hline
& $1_{2}$ & $1_{3}$ & $1_{4}$ & $1_{5}$ & $1_{6}$ & $1_{7}$ & $1_{8}$ & $1_{9}$ \\
\hline
 $1_{2}$ & $1_{3}$ & $\bf{1_{1}}$ & $1_{6}$ & $1_{4}$ & $1_{5}$ & $1_{8}$ & $1_{9}$ & $1_{7}$ \\
 $1_{3}$ & $\bf{1_{1}}$ & $1_{2}$ & $1_{5}$ & $1_{6}$ & $1_{4}$ & $1_{9}$ & $1_{7}$ & $1_{8}$ \\
  $1_{4}$ & $1_{6}$ & $1_{5}$ & $1_{7}$ & $1_{9}$ & $1_{8}$ & $\bf{1_{1}}$ & $1_{2}$ & $1_{3}$ \\
 $1_{5}$ & $1_{4}$ & $1_{6}$ & $1_{9}$ & $1_{8}$ & $1_{7}$ & $1_{3}$ & $\bf{1_{1}}$ & $1_{2}$ \\
 $1_{6}$ & $1_{5}$ & $1_{4}$ & $1_{8}$ & $1_{7}$ & $1_{9}$ & $1_{2}$ & $1_{3}$ & $\bf{1_{1}}$ \\
  $1_{7}$ & $1_{8}$ & $1_{9}$ & $\bf{1_{1}}$ & $1_{3}$ & $1_{2}$ & $1_{4}$ & $1_{6}$ & $1_{5}$ \\
  $1_{8}$ & $1_{9}$ & $1_{7}$ & $1_{2}$ & $\bf{1_{1}}$ & $1_{3}$ & $1_{6}$ & $1_{5}$ & $1_{4}$ \\
   $1_{9}$ & $1_{7}$ & $1_{8}$ & $1_{3}$ & $1_{2}$ & $\bf{1_{1}}$ & $1_{5}$ & $1_{4}$ & $1_{6}$ \\
   \hline
\end{tabular}
\end{center}
{\caption{Singlet multiplications of $\Delta(27)$ symmetry.}}
\end{table}

\section{Elements of $\delta U^{\prime}$}\label{secA2}
The elements of $\delta  U^{\prime}$ are given below \\
\begin{align*}
\delta U_{11}^{\prime}=&\frac{1}{m_{1}^{\prime}-m_{2}^{\prime}}\big[\frac{1}{2}\big( -(a_{2}+b_{2})sin^{2}\theta cos \theta +  \sqrt{2}a_{1}(cos^{2}\theta - sin^{2}\theta)sin \theta+ \\ \nonumber & \sqrt{2}b_{1}(cos^{2}\theta - sin^{2}\theta)sin \theta \big)\big] \hspace{0.3cm} ,
\end{align*}
\begin{align*}
\delta U_{21}^{\prime}=&\frac{1}{m_{1}^{\prime}-m_{2}^{\prime}}\big[\frac{1}{2\sqrt{2}}\big(-(a_{2}+b_{2})cos^{2}\theta sin \theta + \sqrt{2}a_{1}(cos^{2}\theta - sin^{2}\theta)cos \theta + \nonumber \\ &\sqrt{2}b_{1}(cos^{2}\theta - sin^{2}\theta)cos \theta \big)\big] + \frac{1}{m_{1}^{\prime}-m_{3}^{\prime}}\big[\frac{1}{2\sqrt{2}}\big( \sqrt{2}a_{1}cos \theta - \nonumber \\
& \sqrt{2}b_{1}cos \theta -(a_{2}-b_{2})sin \theta\big)\big] \hspace{0.3cm} ,
\end{align*}
\begin{align*}
\delta U_{31}^{\prime}=&\frac{1}{m_{1}^{\prime}-m_{2}^{\prime}}\big[\frac{1}{2\sqrt{2}}\big(-(a_{2}+b_{2})cos^{2}\theta sin \theta +  \sqrt{2}a_{1}(cos^{2}\theta - sin^{2}\theta)cos \theta + \nonumber \\
&\sqrt{2}b_{1}(cos^{2}\theta - sin^{2}\theta)cos \theta \big)\big] +  \frac{1}{m_{1}^{\prime}-m_{3}^{\prime}}\big[\frac{1}{2\sqrt{2}}\big( -\sqrt{2}a_{1}cos \theta + \nonumber \\
& \sqrt{2}b_{1}cos \theta +(a_{2}-b_{2})sin \theta\big)\big] \hspace{0.3cm} ,
\end{align*}
\begin{align*}
\delta U_{12}^{\prime}= \frac{1}{m_{2}^{\prime}-m_{1}^{\prime}}\big[\frac{1}{2}\big( -(a_{2}+b_{2})cos^{2}\theta sin \theta+\sqrt{2}cos \theta(cos^{2}\theta-sin^{2}\theta)(a_{1}+b_{1})\big)\big] \hspace{0.3cm} ,
\end{align*}
\begin{align*}
\delta U_{22}^{\prime}=&\frac{1}{m_{2}^{\prime}-m_{1}^{\prime}}\big[\frac{1}{2\sqrt{2}}(a_{2}+b_{2})sin^{2}\theta cos \theta- \sqrt{2}sin \theta (cos^{2}\theta - sin^{2}\theta)(a_{1}+b_{1})\big] + \nonumber \\ & \frac{1}{m_{2}^{\prime}-m_{3}^{\prime}}\big[\frac{1}{2\sqrt{2}}\big((a_{2}- b_{2})cos \theta- \sqrt{2}(-a_{1}+b_{1})sin \theta\big)\big] \hspace{0.3cm}  ,
\end{align*}
\begin{align*}
\delta U_{32}^{\prime}=&\frac{1}{m_{2}^{\prime}-m_{1}^{\prime}}\big[\frac{1}{2\sqrt{2}}(a_{2}+b_{2})sin^{2}\theta cos \theta- \sqrt{2}sin \theta(cos^{2}\theta - sin^{2}\theta)(a_{1}+b_{1})\big] + \nonumber \\
& \frac{1}{m_{2}^{\prime}-m_{3}^{\prime}}\big[\frac{1}{2\sqrt{2}}\big((-a_{2}+b_{2})cos \theta+ \sqrt{2}(-a_{1}+b_{1})sin \theta\big)\big]  \hspace{0.3cm} ,
\end{align*}
\begin{align*}
\delta U_{13}^{\prime}=&\frac{1}{m_{3}^{\prime}-m_{1}^{\prime}}\big[\frac{1}{2}\big(\sqrt{2}cos^{2}\theta(-a_{1}+b_{1})+ (a_{2}-b_{2})sin \theta cos \theta\big)\big]+ \nonumber \\
& \frac{1}{m_{3}^{\prime}-m_{2}^{\prime}}\big[\frac{1}{2}\big((-a_{2}+b_{2})sin \theta cos \theta + \sqrt{2}(-a_{1}+b_{1})sin^{2}\theta\big)\big] \hspace{0.3cm} ,
\end{align*}
\begin{align*}
\delta U_{23}^{\prime}=&\frac{1}{m_{3}^{\prime}-m_{1}^{\prime}}\big[\frac{-1}{2\sqrt{2}}\big(\sqrt{2}sin \theta cos \theta(-a_{1}+b_{1})+ (a_{2}-b_{2})sin^{2}\theta \big) \big]+ \nonumber \\
& \frac{1}{m_{3}^{\prime}-m_{2}^{\prime}}\big[\frac{1}{2\sqrt{2}}\big((-a_{2}+b_{2})cos^{2}\theta+ \sqrt{2}(-a_{1}+b_{1})sin \theta cos \theta\big)\big] \hspace{0.3cm} ,
\end{align*}

\begin{equation}\label{eq4}
\delta U_{33}^{\prime}=\delta U_{23}^{\prime}
\end{equation}
\end{appendices}

\bibliographystyle{unsrt}
\bibliography{wilinahindawi.bib}

\end{document}